\documentclass[12pt,preprint,showkeys,showpacs]{aastex}
\usepackage{graphicx}
\usepackage{amsmath}
\begin{document}
%\draft
%\preprint{LA-UR-06-7399}
\title{Systematic quantum effects on screening of fusion rates in white dwarfs}
\author{Shirish M. Chitanvis}
\affil{
Theoretical Division,
 Los Alamos National Laboratory,
 Los Alamos, New Mexico 87545}
 \email{shirish@lanl.gov}
\date{\today}
\begin{abstract}
Electron degeneracy effects are dominant in ultra-dense plasmas (UDP), such as those found in white dwarfs. These effects can be treated systematically by obtaining an expansion of the screening length in inverse powers of $\hbar^{2}$. The theory exhibits Thomas-Fermi-like screening in an appropriate regime.  In general, our theory leads to an ${\cal O}(1)$ effect on the enhancement of fusion rates in white dwarfs. Further, it is shown analytically for these stellar conditions that Bose statistics of nuclei have a negligible effect on the screening length, in consonance with Monte Carlo simulations found in literature.
\end{abstract}

\maketitle

\newpage
\section{Introduction}

The evolution of a white dwarf towards a black hole is marked by fusion reactions between carbon  and oxygen nuclei.  These reactions are enhanced by the screening of the Coulomb interaction between fusing nuclei by the surrounding plasma. Many papers have estimated this enhancement factor\citep{salpeter61-1,salpeter69,janovici77,slattery82,itoh90,ogata97,ichimaru99,pollock04,comparat05,gasques05}. It is difficult to gauge the accuracy of these calculations. Our goal here is to provide a systematic basis for the estimation of the fusion enhancement factor in white dwarfs.

Nuclei in white dwarfs are submerged primarily in a sea of electrons.  This is because the probability of finding a nucleus, say, in a two-component UDP (one specie of nuclei of charge $Ze$, and electrons) is down by a factor of $1/Z$, compared to finding an electron. Furthermore, Coulomb repulsion between nuclei is stronger than the thermal energy, so that nuclei tend to remain separated over a relatively large distance. Consequently, we expect the screening to be dominated by a sea of electrons. These electrons are degenerate since the associated fermionic chemical potential is much larger than the thermal energy.

Our earlier paper\citep{chitanvis06-3} was directed towards understanding the screening effect of electronic quantum fluctuations on fusion reactions near the center of the sun\citep{salpeter54,gruzinov98}.  That paper showed quantum effects are negligible, via a systematic expansion of the screening length in powers of $\hbar^{2}$, putting to rest a controversy that has ebbed and flowed over the years\citep{bahcall02}. In this paper we consider a plasma where quantum effects dominate, which is quite opposite to the solar plasma. And so we contemplate the theory in our previous paper in inverse powers of $\hbar^{2}$, which will apply when the effects of electron degeneracy dominate.

This paper provides an analytical underpinning to the numerical techniques used in the past to study fusion in UDP\citep{ogata97,pollock04}. These papers show that the effect of indistinguishability of nuclei on the screening of fusion rates is small. Further, our technique provides an alternative, systematic method for the calculation of the enhancement factors of fusion rates in UDP. Our estimates of enhancement are much more conservative than those given by Ichimaru and Kitamura\citep{ichimaru99}, and in general agreement with recent results of Gasques et al\citep{gasques05}.

\section{Screening Formalism for degenerate electrons}

Let us begin with the classical Poisson-Boltzmann equation for a single species of ions and electrons:

\begin{eqnarray}
 -\nabla^2 \phi &&= 4 \pi \rho \nonumber\\
\rho &&=\rho_{+}+\rho_{-}\nonumber\\
\rho_{+} &&=e ~ n~Z ~\exp(-Z e \phi/k_B T) \nonumber\\
\rho_{-}&&=- e~n~Z ~\exp(e \phi/k_B T) 
\label{pb1}
\end{eqnarray}
where $e$ is the magnitude of the electronic charge, $k_{B}$ is Boltzmann's constant, 
$n$ is the average number density,
$Z e$ is the ionic charge,
and $T$ is the temperature of the system.
We shall work in the linear regime, by retaining only terms first order in $\phi$, leaving to the next section a discussion of nonlinear terms resulting from the Boltzmann distribution:

\begin{eqnarray}
\nabla^2 \phi &&\approx \left(\frac{4 \pi n (Z^{2}n +Z n)e^{2}}{k_{B}T}\right) ~\phi\nonumber\\
&&\equiv \Lambda_{0}^{-2} \phi \nonumber\\
\Lambda_{0} &&= \sqrt {\frac{k_{B}T}{4 \pi n e^{2}(Z^{2}+Z)}}
\label{pb2}
\end{eqnarray}
where $\Lambda_{0}$ is the classical screening length.

We shall now generalize this method to one where nuclei are treated classically, but electrons are treated quantum mechanically\citep{chitanvis06-3}. 
The quantum-mechanical version of the linearized Poisson-Boltzmann equation for a single species of ions and electrons may be written in analogy with Eqn. \ref{pb1}:

\begin{eqnarray}
 -\nabla^2 \phi &&= 4 \pi \rho \nonumber\\
\rho &&=\rho_{+}+\rho_{-}\nonumber\\
\rho_{+} &&\approx  e ~ n~Z ~\left(1-\frac{Z e \phi}{k_{B}T}\right)\nonumber\\
\rho_{-}&&=- e~\vert \psi(\{\vec r\})\vert^{2}
\label{qpb1}
\end{eqnarray}
where $\psi$ is the many-body quantum wave-function for electrons, and $\{\vec r\}$ refers collectively to the electrons in the system, and $\phi$ is the electrostatic potential.

We now invoke the following scaled variables, in order to ease subsequent calculations:

\begin{eqnarray}
\tilde \phi &&= Z e \phi/k_{B}T \nonumber\\
\tilde \psi&&= \Lambda^{3/2}\psi\nonumber\\
\Lambda &&= \sqrt{\frac{k_B T}{4 \pi Z^{2}n e^2}}\nonumber\\
\vec r' && = \frac{\vec r}{\Lambda} \nonumber\\
\Gamma' &&= \frac{Z e^{2}}{\Lambda k_{B}T}
\label{scl}
\end{eqnarray}
where $\Gamma'$ is defined differently from the usual plasma parameter. Notice also that the scalar potential has been scaled differently than in our previous paper\citep{chitanvis06-3}.  This is done to allow for a convenient analysis of higher order contributions in the next section.
Note that the first of Eqns.\ref{scl} shows that we are using $k_{B}T$ as the energy scale.
The electrostatic potential is then given by:

\begin{equation}
 \nabla'^2 \tilde \phi = (\tilde \phi + 4 \pi \Gamma' \vert\tilde \psi \vert^{2} - 1) 
\label{sc2}
\end{equation}

This equation may be obtained from a Lagrangian density:

\begin{eqnarray}
{\cal L}_{0}&&= -\frac{1}{2} \vert \vec \nabla \tilde \phi\vert^{2}- v(\tilde \phi, \tilde \psi)\nonumber\\
v(\tilde \phi, \tilde \psi) &&= \frac{1}{2}\tilde \phi^{2} + 4 \pi \tilde \phi\Gamma' \vert\tilde \psi \vert^{2} -\tilde \phi
\label{sc3}
\end{eqnarray}

The corresponding Hamiltonian density can be easily derived:
\begin{equation}
{\cal H}_{0}= \frac{1}{2} \vert \vec \nabla \tilde \phi\vert^{2}+ v(\tilde \phi, \tilde \psi)
\label{sc3a}
\end{equation}

We will now introduce second-quantized notation to deal with the statistics of electrons:

\begin{equation}
v(\tilde \phi, \tilde \psi)  \to v(\tilde \phi, \tilde \psi_{\pm}) = \frac{1}{2}\tilde \phi^{2}- \tilde \phi + 4 \pi \tilde \phi\Gamma' (\tilde \psi^{\dagger}_{+}\tilde \psi_{+}+\tilde \psi^{\dagger}_{-}\tilde \psi_{-})
\label{sc3b}
\end{equation}
where $\tilde \psi_{\pm}$ are Grassmann variables, and the subscripts refer to the spin of the electrons.
The co-existence of Grassmann variables and scalars in Eqn.\ref{sc3b} is not problematic, since we shall use this discussion solely to define a partition function for the entire system. And soon thereafter we shall integrate over the electron degrees of freedom, so that only a functional involving the scalar potential survives.

The total Hamiltonian ${\cal H}$ for the system, including the quantum-mechanical part for the electrons is:

\begin{eqnarray}
{\cal H} &&= {\cal H}_{0} + {\cal H}_{Q} \nonumber\\
{\cal H}_{Q} &&= \Delta_{Q}(m) (\vert \vec \nabla \tilde \psi_{+}\vert^{2}+\vert \vec \nabla \tilde \psi_{-}\vert^{2} )
\label{sc4}
\end{eqnarray}

The quantum correction has been encapsulated in the following dimensionless parameter:

\begin{equation}
\Delta_{Q}(m)=\left( \frac{\hbar^{2}\Lambda^{-2}}{2 m k_{B}T}\right)\label{qmech1}
\end{equation}
where $m$ is the mass of the electron, $e$ its electronic charge, and $a_{0}$ is the Bohr radius.

Since the temperatures we consider are $T\sim {\cal O}(10^{8}  K) \sim {\cal O}(10 keV)$, the mass density $\rho\sim 10^{9} g~cm^{-3}$, and the rest energy of the electron is $0.55 MeV$, it follows that the non-relativistic approximation employed in Eqn.\ref{sc4} is valid.

The partition function may be written in scaled variables as:

\begin{equation}
{\cal Z} = \int~{\cal D}\tilde\phi~{\cal D}^{2}\tilde \psi_{\pm}~\exp(- \int d^{3}x' ({\cal H}_{0}+{\cal H}_{Q}))\label{pf1}
\end{equation}
where it is understood that $k_{B}T=1$ in the units we are using.

Note that the total number of electrons associated with each ion of charge $Ze$ is $Z$, and is obtained via  $<(\tilde \psi_{+}^{\dagger}\tilde \psi_{+}+\tilde \psi_{-}^{\dagger}\tilde \psi_{-})>=Z n$ (where the angular brackets indicate an expectation value). We will indicate shortly how one may impose this constraint upon the system using a Lagrange multiplier. In condensed matter physics, this is done via the electronic chemical potential.  In our problem, it will turn out to be more convenient to institute this constraint via a functional involving just the electrostatic potential.
The quadratic nature of the energy functional in Eqn.\ref{pf1} allows us to perform the functional integration over the Grassmann variables associated with the electronic degrees of freedom\citep{ramond}, allowing us to obtain:

\begin{eqnarray}
{\cal Z} &&\sim \int {\cal D}\tilde \phi ~\exp(-\int d^{3}x' ((1/2)\vert \vec \nabla \tilde \phi\vert^{2}+(1/2) \tilde\phi^{2} - \tilde \phi))~ {\rm {Det}}({\cal F}) \nonumber\\
{\rm {Det}}({\cal F})&& = \exp (Tr \ln({\cal F}))\nonumber\\
{\cal F} && \equiv -\Delta_{Q}(m)~\nabla^{2 } + 4 \pi \Gamma' \tilde \phi
\label{eff1}
\end{eqnarray}

Having integrated over the electronic degrees of freedom, we are left with an effective energy density in terms of the electrostatic potential alone.  We could have also done things the other way, integrating over the electrostatic potential in the partition function, leaving a quartic in the fermionic variables, as is conventionally done. Our procedure can be said to have {\em bosonized} our plasma, since we now only have the scalar potential to investigate.  We shall show below how our method leads to useful insights into the statistics of the plasma.

One way to impose charge conservation, which was discussed just below Eqn.\ref{pf1}, via a Lagrange multiplier by making the following addition ($\Delta {\cal H}$) to the energy density:

\begin{equation}
\Delta {\cal H} =  4 \pi \nu \tilde \phi
\label{lmul}
\end{equation}

Here $\nu$ may be interpreted physically as a uniform charge density, which will be adjusted to ensure either overall charge neutrality.
We now need to evaluate the determinant of the operator obtained in the process of performing the quadratic functional integral over fermionic variables. This is conveniently performed in Fourier space.

\begin{equation}
Tr \ln({\cal F}) \equiv \int ~\frac{d^{3}k}{(2 \pi)^{3}}~ \ln (4 \pi \Gamma' \hat \phi(k) + \Delta_{Q}(m)~k^{2})
\label{eff2}
\end{equation}
where $\hat \phi(k)$ is the Fourier transform of $\tilde \phi$.

For temperatures in the range of $2\times10^{8} K$, and using a number density for our plasma $\sim 2\times 10^{32}cm^{-3}$ (which corresponds to a mass density of $3\times10^{9}g-cm^{-3}$), we find with $Z\sim10$, it turns out that $\Lambda \sim1 ~Fermi (10^{-12}cm)$, $\Gamma' \sim 60$, and $\Delta_{Q}(m)  \sim 0.3\times10^{5}$.  This is an indication that quantum corrections are dominant in this system.  Hence $4 \pi \Gamma' /\Delta_{Q}\sim 2\times10^{-2} < < 1$ is an excellent choice for a perturbation parameter:

\begin{equation}
\ln (4 \pi \Gamma' \hat \phi(k) + \Delta_{Q}~k^{2}) \approx \ln( \Delta_{Q}~k^{2}) + \frac{4 \pi \Gamma' \hat \phi(k)}{\Delta_{Q}~k^{2}}-\left( \frac{4 \pi \Gamma' \hat \phi(k)}{\Delta_{Q}~k^{2}}\right)^{2}~+~{\cal O}(({4 \pi \Gamma' }/{\Delta_{Q}})^{3})
\label{q2}
\end{equation}

Terms devoid of field variables in Eqn. \ref{q2} will be ignored. For the particular case of a dense UDP considered here, the cubic term ignored in the above expansion is of ${\cal O}(10^{-2})$ compared to the bilinear terms we retained. Thus, higher order terms can be safely ignored.
We must also consider for consistency higher order terms in the expansion of the Boltzmann distribution corresponding to the nuclear charges. We shall do so in the next section.

Note the appearance of powers of $k^{-2}$ in Eqn.\ref{q2}.  This could be problematic when we go to a representation in real space, as Fourier Transforms of $k^{-2p}$, with $p=2,3,...$ do not exist.  For this reason, we make the following ansatz for $\hat \phi(k)$ which resolves this issue entirely:

\begin{eqnarray}
\hat \phi (k) &&= k^{2} \chi(k) \nonumber\\
\tilde \phi(\vec r) &&= -\nabla^{2} \chi(\vec r)
\label{ren1}
\end{eqnarray}

Then the corresponding Lagrangian density in real space may be written to order $\Delta_{Q}^{-2}$ as follows:

\begin{eqnarray}
{\cal L}_{effective} &&= -\frac{1}{2}\vert \vec \nabla \nabla^{2}\chi(\vec r)\vert^{2}\nonumber\\
&& -(1+4 \pi \nu) \nabla^{2} \chi(\vec r)+\frac{1}{2}(\nabla^{2} \chi(\vec r))^{2}-V_{{effective}}(\chi(\vec r))\nonumber\\
V_{{effective}}(\chi(\vec r))&&\approx  b  \chi(\vec r) + b^{2}~ \chi(\vec r)^{2}
\label{qc3}
\end{eqnarray} 
where $b = 4 \pi \Gamma/\Delta_{Q} $.
$\nu$ will be chosen to zero out the coefficient of the term linear in $\chi$, in order to maintain overall charge neutrality.
The  equation of motion obtained by extremizing the above Lagrangian with respect to variations in the field $\chi$ is linear, albeit of order six:

\begin{eqnarray}
(\Delta^{3} -  \Delta + b^{2}) \chi(\vec r) &&=-b \nonumber\\
\Delta &&\equiv \nabla^{2}
\label{qc4}
\end{eqnarray}

The particular solution of the inhomogeneous solution of Eqn.\ref{qc4} turns out to be, by inspection, simply $-b^{-1}$. This constant is physically meaningless, as it does not contribute to the scalar potential and hence to the charge density. To this particular solution must be added the solution of the homogeneous part of the differential equation.
Upon factorizing the trinomial above, the homogeneous part of the differential equation may be cast as:

\begin{equation}
(\Delta - s_{3}) (\Delta - s_{2}) (\Delta - s_{1}) \chi(\vec r) = 0
\label{qc5}
\end{equation}
where $s_{1},s_{2},s_{3}$ are the roots of the trinomial in Eqn.\ref{qc4}. These can be found can be done most easily using Mathematica.  Upon careful examination of these roots, only one yields the correct limit for the screening length given by $1/\sqrt s_{1},1/\sqrt s_{2},1/\sqrt s_{3}$ in the limit that $\hbar \to \infty$. The screening length must go to infinity in this limit due to quantum fluctuations.-- One must remember that quantum fluctuations increase the screening length\citep{chitanvis06-3}. We will denote this root by $s_{1}$.  Its full expression is uninformative. However, in limiting cases it may be written down simply.

For $\hbar \to \infty$,

\begin{equation}
s_{1} \approx  b <<1
\label{r1}
\end{equation}

For this limiting case, the dimensionless screening length is then $1/\sqrt b >>1$.

Hence we shall choose the physically interesting solution to satisfy:

\begin{equation}
(\nabla^{2}- s_{1})\chi(\vec r) = 0
\label{qc5a}
\end{equation}
and this will be sufficient to guarantee that the sixth order differential equation is automatically satisfied.
One may verify the screening length argument by noting that in one dimension $\exp(-x/\sqrt b)$ solves the differential equation in the limit that $\hbar \to \infty$. We have used here a finite-temperature formalism involving not only an electron gas, but also an admixture of positive nuclei to study screening in the limit that electronic quantum fluctuations dominate.  We therefore expect to get for $\hbar \to \infty$ a screening length which is similar, but not identical to the Thomas-Fermi length, which is obtained from a zero-temperature formalism.  As given in Fetter and Walecka\citep{fetter-walecka}, the Thomas Fermi screening length for an electron gas is given by:

\begin{eqnarray}
\lambda_{Thomas-Fermi} &&\approx  \sqrt{\frac{a_{0} r_{0}}{2.95}}\nonumber\\
a_{0}&&= \frac{\hbar^{2}e^{2}}{2 m}
\label{tf1}
\end{eqnarray}
where $a_{0}$ is the Bohr radius, and $r_{0}$ is the mean free distance between charges.

In our case, the screening length $1/\sqrt b$ ($\hbar\to \infty$) turns out to be:

\begin{eqnarray}
\lambda_{screening} &&\approx \sqrt{\frac{a_{0} \ell_{0}(T)}{4}}~\propto \hbar \nonumber\\
\ell_{0}(T) &&= \frac{\Lambda(T)}{4 \pi}\
\label{tf2}
\end{eqnarray}
where the temperature-dependent scale $\ell_{0}(T)$ plays the same role as does $r_{0}$ in Eqn.\ref{tf1}.
Notice that the screening length displays a rather weak $\sim T^{1/4}$ temperature dependence, in qualitative agreement with the conclusions of Ichimaru\citep{ichimaru99} for electronic screening in UDP.
In general, the screening length is given by $1/\sqrt s_{1}$ and will be discussed in much detail in of section 4. 

\section{Non-linear screening effects of nuclear charges}

In the previous section, we discussed in detail how non-linear terms arising from electron degeneracy are safely of much smaller magnitude than those retained.  The issue appears to get turned around when one considers non-linear terms arising from higher order terms in the expansion of the Boltzmann distribution corresponding to nuclear charges.  This is because the coefficients of the nonlinear terms are larger than those found for degenerate electrons in the previous section. The characteristic Coulomb energy $Z^{2}e^{2}/r_{0}>>  kT$, for $Z=6$. Hence we are in the regime of strong screening\citep{salpeter54}. What this means is that the positive charges remain separated by a large distance, with electrons performing most of the screening of nuclear charges. We will see shortly how this is borne out somewhat more rigorously.

The issue may be compactly discussed by noting that retention of the quadratic term in Eqn. \ref{qpb1} leads to the following cubic modification of Eqn.\ref{sc3b}:

\begin{equation}
v \to v - \frac{1}{6} \tilde \phi^{3}
\label{vf1}
\end{equation}

The corresponding partition function for our classical system resembles one for a Euclidean scalar quantum field theory\citep{ramond}. Effects of the non-linear terms can be evaluated perturbatively, using Feynman diagrams.
Feynman diagrams can be used to estimate the leading order contribution from this cubic term to the self-energy of the system (or, equivalently, the dielectric constant, or the screening length). It turns out that the actual value of the contribution is numerically quite small.  This suggests that the linear screening approximation we retained in the previous section is an acceptable approximation.

In order to perform this calculation, we formally ignored terms of ${\cal O}(b)\sim 10^{-2}<<(1/6)$ in our diagrammatics.  We then used Mathematica to formulate symbolically the lowest order contributions from the cubic potential. The term that survives is a polarization-like diagram which comes from terms of ${\cal O}((1/6)^{2})$. The calculation is done as usual in momentum space. At this point the momentum variable is simply set to zero, so that we get an expression for the inverse square of the screening length:

\begin{eqnarray}
&&\hat \Sigma(k)_{polarization-like}(\vec k)  = \frac{1}{12} \int \frac{d^3p}{(2 \pi)^3} \hat G_0 (\vec k) \hat G_0(\vec p - \vec k) \nonumber\\
&&\hat G_0(\vec p) = \frac{1}{p^{2} + 1}
\label{dh5a}
\end{eqnarray}
where $\hat \Sigma(k)$ is the usual self-energy.  It can be easily related to the dielectric constant of the UDP. We shall restrict attention to $k=0$, when the self energy it reduces to the inverse square of the screening length, and is sufficient to allow us to gauge its magnitude relative to the degenerate contribution in the previous section.

Whence:
\begin{eqnarray}
\Lambda_{polarization-like}^{-2}\equiv\hat \Sigma(k=0)_{polarization-like}  &&=\frac{1}{96 \pi}~\nonumber\\
&&\sim 3.3\times10^{-3}~{\rm(white~ dwarfs)}\nonumber\\
\Lambda_{polarization-like} &&\sim 1.7 \times 10^{-11}cm ~{\rm(white~ dwarfs)}
\label{dh5aa}
\end{eqnarray}

We see that the screening length from the cubic term in the energy functional (due to nuclear charges alone)  is larger by an order of magnitude than that obtained in the previous section by an order of magnitude ($\Lambda_{screening} \sim 10^{-12}cm$ for a white dwarf, from Eqn.\ref{qc5}). This difference in the screening due to nuclear and electronic charges means that electrons can be in closer proximity to a nucleus than another another nucleus. 

The screening lengths appears as its inverse square in the corresponding differential equation for the scalar potential. Hence one can combine the contributions from the previous section, with the current one:

\begin{equation}
(\nabla^{2} - \Lambda_{polarization-like}^{-2} - \Lambda_{screening}^{-2})\tilde \phi(\vec r) =0
\label{e1}
\end{equation}

Thus the lowest order contribution of the cubic term to the screening length in Eqn.\ref{e1} is about two orders of magnitude smaller, compared to the screening contribution obtained in the previous section. We shall therefore ignore this non-linear correction.

The conclusions of this section may have to be modified if  higher order terms turn out to be larger.
For that purpose, we retain the quartic term in the energy functional arising from the Boltzmann distribution for the nuclear charges, so that Eqn.\ref{vf1} is modified as follows: 

\begin{equation}
v \to v - \frac{1}{3!} \tilde \phi^{3} + \frac{1}{4!} \tilde \phi^{4}
\label{vf2}
\end{equation}

The lowest order contribution from the quartic term is the setting-sun diagram\citep{ramond}.   There are no cross-terms at this order between the cubic and quartic terms.

\begin{equation}
\Sigma_{setting-sun}(p) = \frac{1}{6} \int \int \frac{d^3k_1}{(2 \pi)^3} \frac{d^3k_2}{(2 \pi)^3} \hat G_0 (k_1) \hat G_0(k_2) \hat G_0(|\vec p - \vec k_1 - \vec k_2|)\label{dh5b}
\end{equation}

Using $t^{-1} = \int d\lambda \exp(-\lambda t)$, converting the momentum integrals to center of mass and relative co-ordinates, and using dimensional regularization, we can compute the self-energy for $p=0$ in the following form:

\begin{eqnarray}
\Sigma_{setting-sun}(0) &&= \left(\frac{1}{6}\right)~ \frac{L ~ I}{8 \pi^4} \nonumber\\
L &&= \int_0^\infty \frac{\exp(-\xi x)}{x^{1-\epsilon}} dx\nonumber\\
I  &&= \int_0^\infty \frac{\ln(\xi)-\ln(1+y)}{(1+4 y)^{1+\epsilon}} dy 
\label{dh5c}
\end{eqnarray}
where $\xi \to 0^+$, a small-distance cut-off, and $\epsilon \to 0^+$ have been inserted to guarantee convergence.
Using Mathematica, it can be shown that:

\begin{eqnarray}
L &&= ({\xi})^{-\epsilon} \Gamma(\epsilon) \nonumber\\
I &&= \frac{\ln(\xi)}{4 \epsilon}-\frac{1}{16} \left( \frac{3^{-\epsilon} ~4~\rm{cosec}(\pi \epsilon)}{\epsilon} + \Gamma(-\epsilon)~ _2F_1(1,1;2-\epsilon;1/4)/\Gamma(2-\epsilon) \right)
\label{dh5d}
\end{eqnarray}
where $_2F_1(a,b;c;z)$ is the hypergeometric function.

Employing Laurent expansions within Mathematica in powers of $\epsilon$, and upon using counter-terms that account for divergences, one obtains
the following correction to the square of the screening length, correct to lowest order in the quartic term:

\begin{eqnarray}
&&\Sigma_{setting-sun}(0)   =\nonumber\\
&&\frac{1}{48 \pi^{4}}\left(-\gamma-\log (\xi )+\frac{1}{48} \left(-2 \pi ^2+12 \gamma  \log
   \left(\frac{4}{3}\right)\right)-\frac{1}{16}~ \left(\left(2 \log
   ^2(3)-\gamma_{1}\right)\right) \right)\nonumber\\
&&\sqrt \xi = \frac{a }{\Lambda}
\label{dh6}
\end{eqnarray}
where $a<<\Lambda$ is a microscopic length cut-off required to render the integrals finite, $\gamma\approx 0.577$ is Euler's constant, and $\gamma_{1} \approx 0.572$ is the value of the derivative of the Hypergeometric function which appears in Eqn.\ref{dh5d}. A logarithmic dependence of our answer on this cut-off implies a relative insensitivity to this parameter. It is clear that the theory used in this section is certainly not valid at the nuclear level, and so we will use $\xi=a /\Lambda\sim 10^{2}$.
A cursory examination shows that this particular diagram yields a small contribution ${\cal O}(10^{-4})$ contribution to the self-energy.  Then following the argument above, for the cubic term, this screening contribution can be ignored as well. Thus there appears to be a trend for the classical, nonlinear terms to be small.

Of course, further issues may arise in this perturbative argument as even higher order terms arising from the Boltzmann distribution are contemplated. We will leave these questions for future investigation. It may be possible to extend to white dwarfs the methods utilized by Brown et al\citep{brown06} for obtaining the screening length for a dilute, highly charged plasma.

\section{Quantum effects of nuclei on screening}

Over the years, researchers have delved into the importance of applying a quantum-statistical treatment to the nuclei surrounding the ones undergoing fusion. This is a reasonable point to investigate, given the extremely high densities available in white dwarfs. Path integral Monte Carlo  (PIMC) techniques have led to the discovery that the bosonic nature of nuclei\citep{itoh90,ogata97,pollock04} makes a calculable, small contribution to the screening length, or equivalently, to the dielectric constant of the UDP under consideration.  Here we provide an analytical underpinning to that observation. The second point that needs to be reinforced is that quantum effects of nuclei in the UDP are small in general, compared to the electronic contribution.  The argument is basically that of Born and Oppenheimer, who showed that the nuclear mass is so large compared to the mass of an electron that an adiabatic approximation can be applied. That argument has to be extended to finite temperatures.

We begin by considering a UDP consisting of spin-zero nuclei, in addition to a sea of degenerate electrons.
Thus the Poisson-Boltzmann equation (Eqn.\ref{sc3b}) will be replaced by:

\begin{equation}
v(\tilde \phi, \tilde \psi)  \to v_{B}(\tilde \phi, \tilde \Psi,\tilde \psi_{\pm}) =4 \pi Z \tilde \phi\Gamma (\tilde \Psi^{\dagger}~\tilde \Psi)  + 4 \pi \tilde \phi\Gamma (\tilde \psi^{\dagger}_{+}\tilde \psi_{+}+\tilde \psi^{\dagger}_{-}\tilde \psi_{-})
\label{sc3bb}
\end{equation}
where $\tilde \Psi^{\dagger}$, $\tilde \Psi$ are the second-quantized creation and annihilation operators corresponding to bosonic nuclei,
and we have continued to assume that the photons are numerous and hot that they can be treated classically.

Equation \ref{sc4} must be modified to account for the free Hamiltonian of the nuclei:

\begin{eqnarray}
{\cal H}_{Q} &&\to  {\cal H}_{Q} + {\cal H_{N}}_{Q} \nonumber\\
 {\cal H_{N}}_{Q}&&=\Delta_{Q}(M) ~\vert \vec \nabla \tilde \Psi\vert^{2}
\label{sc44}
\end{eqnarray}
where $\Delta_{Q}(M)$ is defined via Eqn.\ref{qmech1}, with the mass of the electron being substituted by the mass of the nucleus under consideration. It has been assumed that the interaction between nuclei are dominated by the Coulomb potential.  This is a reasonable assumption to make in view of the fact that in this dense plasma, the nuclei are separated on the average by about $10 Fermis$, as shown in the previous section.

The corresponding functional integrals involving the nuclear field variables can be done, just as the fermionic degrees of freedom were accounted for.  The net result for the partition function is:

\begin{equation}
{\cal Z} \sim \int {\cal D}\tilde \phi ~\exp(-\int d^{3}x' ~(1/2)\vert \vec \nabla \tilde \phi\vert^{2})~ {\rm {Det}}({\cal F}) ~ {\rm {Det}}({\cal B})^{-1} 
\label{time11}
\end{equation}
where

\begin{equation}
Tr \ln({\cal B}^{-1}) \equiv~ -~\int ~\frac{d^{3}k}{(2 \pi)^{3}}~ \ln (4 \pi Z \Gamma' \hat \phi(k) + \Delta_{Q}(M)~k^{2})
\label{eff22}
\end{equation}
Using parameters representative of a white dwarf, and given that $M \sim 2\times 10^{4} m$, $Z \sim 10$, it turns out that $\Delta_{Q}(M)\sim10^{-4} \Delta_{Q}(m)$, so that $ \Delta_{Q}(M)/4 \pi Z \Gamma  \sim 10^{-3}$.  This clearly shows that quantum effects, including Bose statistics of nuclei in the UDP are negligible in circumstances representative of a white dwarf.  Thus, we may continue to use a classical treatment for nuclei, as was done in the previous section. One can regain from Eqn.\ref{eff22} the Boltzmann approximation using the method outlined by Chitanvis\citep{chitanvis06-3}.  This will yield systematic, extremely small quantum corrections to the Boltzmann approximation in powers of $\hbar^{2}$.  Considering the nuclei as fermions does not change this conclusion. 

Quantum effects of nuclei are much smaller than the quantum effects due to electrons, primarily due to the large mass difference. As such they can be ignored. Our conclusions concerning the effects of indistinguishability are in general agreement with Itoh\citep{itoh90}, Ogata\citep{ogata97} and Pollock and Militzer\citep{pollock04}. These authors did not consider the effects of electron degeneracy on the same footing as the nuclear quantum effects.

\section{Results}

We shall compare our results with those found in published literature. For succinctness, we shall quote the enhancement factors of pycnonuclear reaction rates given in the review article of Ichimaru and Kitamura\citep{ichimaru99} and results by Gasques et al\citep{gasques05}.  

Our approach provides an integrated, first-principles theory of screening effects due to electrons and associated nuclei.
Furthermore, we have derived a systematic way of obtaining screening effects in a UDP.  As such, it is possible for us to estimate the accuracy with which we can calculate our enhancement factors.  In order to do that for white dwarfs, we must first generalize our formalism to a binary ionic mixture (BIM), e.g. a mixture of $^{12}C$, $^{16}O$ nuclei and associated electrons, as representative of the species in a white dwarf. This is easily accomplished via the following substitutions in all our formulae:

\begin{eqnarray}
Z^{2} &&\to \bar Z^{2}= \frac{n_{1}Z_{1}^{2}+n_{2}Z_{2}^{2}}{\bar n}\nonumber\\
n &&\to \bar n = n_{1}+n_{2}
\label{mix}
\end{eqnarray}
where $n_{1}$, $n_{2}$ are the average number densities of each species in the UDP.
These substitutions arise naturally through a re-derivation of our theory for a BIM.
Generalizations to more than two components is straightforward.

Ichimaru uses improvements over a standard procedure to obtain net enhancement factors.  First, the screening length/dielectric constant of an electron gas in a jellium of positive ions is obtained.  Then various sophisticated methods are used separately to obtain screening effects due to nuclei surrounding the moieties undergoing fusion.  Physically reasonable mixture rules are utilized to obtain the overall enhancement of nuclear rates in a UDP, caused by the screening of the nuclear Coulomb repulsion by intervening charges. It is not possible to gauge the accuracy of such calculations. In general we find our estimates for fusion rate enhancement in white dwarfs are much more conservative than those of Ichimaru and Kitamura\citep{ichimaru99}. In an effort to understand this difference, we found that the overall plasma screening length obtained by us (primarily a degenerate electron contribution), and the electronic screening length of Ichimaru and Kitamura agree reasonably well.  It follows that the difference in the enhancement rates must come from their treatment of screening due to the nuclei.  And we showed in section III that our classical treatment of the heavy nuclei gives a very accurate result for the screening correction.
Furthermore, the values of the classical plasma parameter $\Gamma_{classical}=\bar Z^{2}e^{2}/r_{0}kT << 170$ ($r_{0}$ is the mean-free distance between particles) for the cases listed in Table \ref{table1}.  As such there is no concern regarding the UDP being close to a crystallized state.

Gasques et al utilize a re-parameterized version of the enhancement factor obtained by Slattery et al\citep{slattery82}. Our results agree more reasonably with Gasques et al\citep{gasques05} for the following reason. The enhancement factor of Gasques et al is defined in terms of the plasma parameter of a one-compenent plasma.  In our case, we have a binary mixture. However, since we are in a strong-screening regime\citep{salpeter54}, we suggest that our BIM is a one-component plasma composed only of electrons (for purposes of screening).  In any event, it is instructive to compare the enhancement using a classical electron gas to our results.  We find that the classical result gives an enhancement factor which is roughly twice that obtained for our BIM.  We attribute our lower values of enhancement to quantum effects.

An interesting observation to be made is that if we use the enhancement formula of Gasques et al\citep{gasques05} for a one-component classical plasma composed of positive charges, each having a charge $Z e$, with $Z= 6$, there is remarkable agreement with the results of Ichimaru and Kitamura\citep{ichimaru99}. 

The numerical values obtained for our screening length and comparisons to Ichimaru's enhancement factor\citep{ichimaru99} and that of a classical electron gas\citep{gasques05} have been encapsulated for white dwarfs near ignition in Table \ref{table1}. 

%\vspace{1cm}
\begin{table}[h]
\begin{tabular}{llll}
\hline
\textbf{Quantity} &  \textbf{Current theory} & {\citep{ichimaru99}}&{classical electron gas}\\
\hline
\hline
Model&WD1&\\
\hline
Density & $\leftarrow$  & $3.0\times 10^{9}g-cm^{-3}$& $\rightarrow$\\
T (Kelvin) & $\leftarrow$ & $1.8\times 10^{8} K$&$\rightarrow$\\
Composition & $\leftarrow$&$^{12}C$, $e^{-}$& $\rightarrow$\\
\hline
$\Gamma_{enhance}$ & $ 0.57$  & $ 12.09$ & $0.26$\\
\hline
\hline
Model&WD2&\\
\hline
Density & $\leftarrow$   & $9.0\times 10^{9}g-cm^{-3}$&$\rightarrow$ \\
T (Kelvin) & $\leftarrow$  & $1.1\times 10^{8} K$&$\rightarrow$ \\
Composition &$\leftarrow$  &$^{12}C$, $e^{-}$&$\rightarrow$\\
\hline
$\Gamma_{enhance}$ & $ 1.39$  & $ 23.10$ & $0.73$\\
\hline
\hline
Model&WD3&\\
\hline
Density & $\leftarrow$  &$9.0\times 10^{9}g-cm^{-3}$ &$\rightarrow$ \\
T (Kelvin) &$\leftarrow$  & $3.4\times 10^{7} K$ & $\rightarrow$ \\
Composition & $\leftarrow$& $^{12}C$, $e^{-}$ & $\rightarrow$\\
\hline
$\Gamma_{enhance}$ & $ 6.02$  & $ 20.76$&$ 2.5$ \\
\hline
\hline
Model&WD4&\\
\hline
Density & $\leftarrow$  & $9.0\times 10^{9}g-cm^{-3}$ & $\rightarrow$\\
T (Kelvin) & $\leftarrow$ & $1.1\times 10^{8} K$ & $\rightarrow$ \\
Composition &$\leftarrow$ &$^{12}C$ ($75\%$), $^{16}O$ ($25\%$)$e^{-}$ & $\rightarrow$\\
\hline
$\Gamma_{enhance}$ & $ 1.42$  & $ 23.12$ &$0.71$\\
\hline
\hline

\end{tabular}
\caption{Comparison of quantum corrected rate factors for white dwarfs near ignition, between our calculation and two previous calculations\citep{ichimaru99,gasques05}.  The enhancement of the fusion rate is calculated as 
$\exp(\Gamma_{enhance})$, $ \Gamma_{enhance}= e^{2}/(\Lambda_{screening} k_{B}T)$. The quantum-influenced screening length $\Lambda_{screening}$ is defined in dimensionless terms as $1/\sqrt s_{1}$ via Eqn.\ref{qc5}.
The different scenarios for white dwarfs, viz., models $WD1-WD4$ are described in Ichimaru's paper\citep{ichimaru99}.}
\label{table1}
\end{table}
%\vspace{1cm}

\section{Acknowledgments}

This work was carried out under the auspices of the National Nuclear Security Administration of the U.S. Department of Energy at Los Alamos National Laboratory under Contract No. DE-AC52-06NA25396.

\bibliographystyle{apsrev.bst}

\end{document}